\documentclass[pra,aps,12pt,onecolumn,tightenlines,showpacs,amsmath,amssymb]{revtex4-1}

\usepackage[english]{babel}
\usepackage{afterpage}
\usepackage{epsfig,graphicx,amsmath}
\usepackage{amssymb,bm}
\usepackage{enumitem}
\usepackage{color}
\usepackage{slashed}
\usepackage{latexsym}

\newcommand{\dd}{\mathrm{d}}  
\newcommand{\ii}{\mathrm{i}}  
\newcommand{\ee}{\mathrm{e}}  
\newcommand{\me}{m_\mathrm{e}} 

\newcommand{\tsprvg}{{\tilde K}_{\mathrm V}}

\newcommand{\krkh}{K_{\rm KH}}
\newcommand{\krkho}{K^{(0)}_{\rm KH}}
\newcommand{\tkrkho}{{\tilde K}^{(0)}_{\rm KH}}

\begin{document}

\title{Coulomb-Corrected Strong Field Approximation without Singularities and Branch Points\footnote{This work was presented at the 27th Annual International Laser Physics Workshop (Nottingham, July 16-20, 2018).}}
\author{J Z Kami\'nski}
\email[E-mail address:\;]{Jerzy.Kaminski@fuw.edu.pl}
\author{F Cajiao V\'elez}
\email[E-mail address:\;]{Felipe.Cajiao-Velez@fuw.edu.pl}
\author{K Krajewska}
\email[E-mail address:\;]{Katarzyna.Krajewska@fuw.edu.pl}
\affiliation{Institute of Theoretical Physics, Faculty of Physics, University of Warsaw, Pasteura 5,
02-093 Warsaw, Poland
}
\date{\today}

\begin{abstract}
The domain of validity of the Coulomb-Corrected Strong Field Approximation (CCSFA) is going to be analyzed in relation to the semi-classical dynamics of electrons during 
ionization of hydrogen-like targets. Our analysis is limited to ionization driven by Ti-sapphire laser pulses with intensities up to roughly $10^{14}$\,Wcm$^{-2}$. 
For such parameters, the effects related to radiation pressure are small and the laser field can be described in the dipole approximation. By applying the Magnus expansion for the exact retarded electron propagator we obtain an {\it effective action} which is free of Coulomb singularities and branch points when 
the complex-time trajectories are used. Furthermore, we show that the classical action is exactly recovered as the asymptotic limit of its effective counterpart. 
The applicability of such limit is also discussed. 
\end{abstract}

\maketitle

\section{Introduction}

The great English mathematician and philosopher Bertrand Russell (1872-1970) has once written that `\textit{all exact science is dominated by the idea of approximation}.' 
One could continue his line of thought by saying that \textit{the most beautiful aspects of all exact science are the approximations}. It has to be remembered, however, 
that any approximation has its domain of validity, which is inherently related to it and equally important in its applications. The strong-field physics (SFP) 
is, in fact, one of the best illustrations of the above statements. Recent developments in this field have led to the creation of new branch of modern science, the
\textit{attophysics}~\cite{BK,IK}. One of the most prominent tools in the SFP, the Strong-Field Approximation (SFA) in photoionization, was originally introduced by 
Keldysh~\cite{Keldysh} in the length gauge and further developed by Faisal~\cite{Faisal} and Reiss~\cite{Reiss} for other forms of the Schr\"odinger equation (i.e., in 
the Kramers-Henneberger frame~\cite{kh1,kh2} or in the velocity gauge, respectively). The common feature of those approaches is the approximation of the exact 
scattering state of the photoelectron by the Volkov solution~\cite{Volkov,EJK1998}. Subsequently, such approximation has been extended to treat the non-relativistic 
laser-assisted scattering processes~\cite{Fedorov1} and its relativistic counterpart~\cite{Fedorov2}, where the relativistic form of the Volkov 
solution~\cite{EKK2009,DiPiazza2012,DiPiazza2018} is fully exploited.

Since the SFA does not offer a proper explanation of various experimental results, it requires further developments. For instance, an attempt to incorporate 
the interaction of photoelectrons with the parent ion by means of the Coulomb-Volkov state was undertaken in~\cite{Tsoar,Kornev,Duchateau}. This procedure was 
successfully applied to the analysis of ionization driven by elliptically polarized laser fields~\cite{Ferrante}. (Note that similar investigations were 
recently presented for bi-circular laser fields~\cite{Milo2018}.) Further generalizations, which fully account for the low-frequency approximation, were considered 
in Refs.~\cite{K1} and~\cite{K2} for the scattering and ionization processes, respectively. This approach, called the Coulomb-Volkov Strong-Field Approximation (CVSFA), 
was recently generalized in~\cite{PU2018} by incorporating the density functional theory for the initial bound state of many-electron atoms, or in~\cite{ZZ2018} 
by applying the parabolic quasi-Sturmian-Floquet approach. Also, the scattering states in the high-frequency approximation were explored~\cite{Gav1,Gav2}. 
An alternative approach to the CVSFA, which we call the Born-Series Strong-Field Approximation (BSSFA), is to apply the Born expansion to the final 
electron scattering state in both the binding potential and the laser field. Here, the first two terms of the Born series were successfully used in studies 
of the re-scattering process in ionization~\cite{Milo2018,Resc1}.

Another path of theoretical explorations is related to the eikonal approximation~\cite{eik1,eik2,eik3} and its generalizations~\cite{eik4,KK2015,VKK2016}. In our 
further discussion, we shall refer to those approaches as the Coulomb-Corrected Strong-Field Approximation (CCSFA)~\cite{ccsfa1,ccsfa2,ccsfa3,ccsfa4}.  
In connection to these theoretical investigations is the purely classical analysis of laser-assisted atomic or molecular processes~\cite{classdyn} such as 
ionization~\cite{classion1,classion2}, plasma dynamics in atomic clusters~\cite{classion2a}, and re-collisions~\cite{classresc}. In fact, those works were stimulated 
by the successful application of the classical dynamics to ionization of Rydberg states by microwave fields~\cite{classrf1,classrf2}.

The two groups of approaches mentioned above (the CVSFA and BSSFA vs. the CCSFA and the classical analysis) are contradictory when the behavior of the electron wave-packet 
close to the Coulomb singularity is analyzed. For the first group of approximations (and actually for the full quantum treatment; see, e.g.,~\cite{Nurhuda,num1}) 
this singularity does not present any conceptual or numerical problem. In contrast, the classical analysis shows severe numerical problems when the electron trajectories 
approach the singularity (as discussed in~\cite{classion2}). Furthermore, in the CCSFA, when the method of complex-time electron trajectories is applied, the Coulomb 
singularity leads to singular branch points. Such branch points do not exist in the BSSFA or in the exact numerical solution of the Schr\"odinger equation. 
The aim of our investigations is to determine the origin of these problems and to derive the effective CCSFA (related to the generalized eikonal approximation and eikonal 
perturbation theory~\cite{eik4,KK2015,VKK2016}) which is free of Coulomb singularities and branch points.

Note that the CCSFA and the classical 
dynamics follow from the quantum theory in the limit when the Planck constant goes to 0 or the time evolution is very short. However, it must be taken into account 
that $\hbar$ and time have physical dimensions. Thus, in order to provide a sensible physical meaning of this limit, one has to construct a dimensionless parameter 
out of the Plank constant, the time of evolution, and other relevant physical quantities, such that the CCSFA or the classical dynamics are recovered in such limit. 
This is going to be done below for the ionization of a hydrogen-like ion driven by laser fields of moderate intensity. Such condition guarantees that the dipole approximation 
is valid throughout our analysis. For instance, if the Ti-sapphire laser field (wavelength 800~nm and frequency such that $\hbar\omega_{\rm L}=1.55\,\mathrm{eV}$) is considered, 
the laser pulse should not exceed intensities of the order of $10^{14}\,\mathrm{Wcm}^{-2}$. For larger intensities, relativistic effects related to the radiation 
pressure~\cite{press0,KKpress} (see, also~\cite{press1,press2,press3,press4,press5,press6}) and relativistic mass corrections~\cite{CKKvortex} become already visible.

Our analysis below is based on the Magnus expansion, as opposed to the more general eikonal perturbation theory developed in~\cite{eik4,KK2015,VKK2016}. The Magnus 
expansion~\cite{magnus1,magnus2,magnus2a,magnus3} allows one to construct an approximate exponential representation of the propagator of the system. 
Every order of such expansion corresponds to a partial re-summation of infinite terms of an ordinary Born series. Next, we apply this approximation up to the linear 
term with respect to the binding potential and show under which conditions the classical dynamics is restored. This determines the domain of validity 
of the ordinary CCSFA~\cite{ccsfa1,ccsfa2,ccsfa3,ccsfa4} and it allows us to remove the artificial Coulomb singularities and branch points. We show that the latter arise 
from the application of the relevant asymptotic expansion beyond its domain of validity. We demonstrate that the effective CCSFA, in the appropriate limits, 
leads to the exact Born expansion or to the classical (or complex-time) dynamics. Moreover, it appears that the effective action is complex and accounts for 
the electron wave packet spreading in the course of its time evolution. Due to this fact the electron trajectory is also complex even for real times and real
initial/boundary conditions, but such that in the classical limit its imaginary part vanishes. Note that the method presented here and 
in~\cite{eik4,KK2015,VKK2016} allows to incorporate further terms, which are nonlinear with respect to the static binding potential, into the effective action.

\section{General theory and Magnus expansion}
\label{general}
We start with general statements about the non-relativistic quantum-mechanical evolution of an electron in arbitrary electromagnetic fields. 
Such evolution, from an initial time $t'$ up to a final time $t$, is determined by the retarded propagator $K({\bm r},t;{\bm r}', t')$. If the Hamiltonian of 
the system, in terms of the `primed' coordinates, is $H({\bm r}',t')$, the propagator satisfies the differential equation (see, e.g., Ref.~\cite{KK2015})
\begin{equation}
\Big[-\ii\partial_{t'}-H^*({\bm r}',t')\Big]K({\bm r},t;{\bm r}',t')=\ii\delta(t-t')\delta({\bm r}-{\bm r}').
\label{general1}
\end{equation}
As many studies of ionization relate to the momentum or energy distribution of photoelectrons, it is most convenient to analyze the propagator in momentum space. 
For this reason, we introduce its Fourier transform, $\tilde K({\bm p},t;{\bm r}',t')$, calculated over the non-primed coordinate,
\begin{equation}
K({\bm r},t;{\bm r}',t')=\int\frac{\dd^3 p}{(2\pi)^3} \tilde K ({\bm p},t;{\bm r}',t')\ee^{\ii {\bm p}\cdot {\bm r}}.
\label{general4}
\end{equation}
The differential equation for the transformed propagator, $\tilde K ({\bm p},t;{\bm r}',t')$, can be easily determined by inserting Eq.~\eqref{general4} into~\eqref{general1}. 
In doing  so, we find out that
\begin{equation}
\Big[-\ii\partial_{t'}-H^*({\bm r}',t')\Big]\tilde K ({\bm p},t;{\bm r}',t')=\ii\delta(t-t')\ee^{-\ii{\bm p}\cdot{{\bm r}'}},
\label{general6}
\end{equation}
and, as it can be checked, $\tilde K({\bm p},t;{\bm r}',t')$ fulfills the retardation and initial conditions,
\begin{equation}
\tilde K({\bm p},t;{\bm r}',t')=0\quad{\text{for}}\quad t<t'\quad\text{and}\quad\tilde K ({\bm p},t;{\bm r}',t')\xrightarrow[t\rightarrow t'_+]{}\ee^{-\ii {\bm p}\cdot {\bm r}'},
\label{general7}
\end{equation}
respectively.

As mentioned in~\cite{magnus2}, the Magnus expansion finds applications in many subfields of physics including atomic, molecular, 
and particle physics, quantum electrodynamics, etc. Here, we shall use it to solve the homogeneous equation,
\begin{equation}
\Big[-\ii\partial_{t'}-H^*({\bm r}',t')\Big]\tilde K ({\bm p},t;{\bm r}',t')=0,
\label{magnus1}
\end{equation}
while imposing the conditions~\eqref{general7}. This is equivalent to solving the inhomogeneous equation~\eqref{general6}. 
Thus, we look for the solution of~\eqref{magnus1} in the exponential form~\cite{magnus1}
\begin{equation}
\tilde K ({\bm p},t;{\bm r}',t')=\ee^{\ii F({\bm p},t;{\bm r}',t')},
\label{magnus2}
\end{equation}
where $F({\bm p},t;{\bm r}',t')$ is an a priori unknown function. Such function can be expanded in the power series
\begin{equation}
F({\bm p},t;{\bm r}',t')=F^{(0)}({\bm p},t;{\bm r}',t')+\sum_{\ell=1}^\infty \lambda^{\ell} F^{(\ell)}({\bm p},t;{\bm r}',t'),
\label{magnus3}
\end{equation}
where $\lambda$ is a small and real parameter (see below). The challenge, however, is to determine each term $F^{(\ell)}({\bm p},t;{\bm r}',t')$ in the sum~\eqref{magnus3}. 
Below, we illustrate this for the case when $\tilde K ({\bm p},t;{\bm r}',t')$ can also be represented as a Born series.

First, we separate the total Hamiltonian of the system $H({\bm r}',t')$ into two parts, $H({\bm r}',t')=H_0({\bm r}',t')+\lambda H_{\rm I}({\bm r}',t')$, where $H_{\rm I}$ 
describes a relatively weak interaction compared to the main contributing Hamiltonian $H_0$. This is stressed by a small parameter $\lambda$. Next, we also assume that 
the transformed propagator can be represented as a Born series with respect to $\lambda$. Specifically, up to the first order
in the perturbation, we write it down as
\begin{equation}
\tilde K ({\bm p},t;{\bm r}',t')\approx \tilde K^{(0)} ({\bm p},t;{\bm r}',t')+\ii\lambda \tilde K^{(1)}_{\rm B} ({\bm p},t;{\bm r}',t')+\mathcal{O}(\lambda^2),
\label{magnus5}
\end{equation}
where in general $\tilde K^{(\ell)}_{\rm B} ({\bm p},t;{\bm r}',t')$, $\ell=1,2...,$ is the $\ell$-th term of the Born expansion. On the other hand, if the Magnus expansion is to be applied, 
the Fourier transform of the propagator takes the form [see, Eqs.~\eqref{magnus2} and~\eqref{magnus3}]
\begin{align}
\tilde K ({\bm p},t;{\bm r}',t')=\exp{\big[\ii F^{(0)}({\bm p},t;{\bm r}',t')\big]}\exp{\bigg[\ii\sum_{\ell=1}^\infty\lambda^{\ell} F^{(\ell)}({\bm p},t;{\bm r}',t')\bigg]}.
\label{magnus6}
\end{align}
Therefore, up to the first order in $\lambda$, we obtain
\begin{equation} 
\tilde K ({\bm p},t;{\bm r}',t')\approx\exp{\Big[\ii F^{(0)}({\bm p},t;{\bm r}',t')\Big]}\times\Big(1+\ii\lambda F^{(1)}({\bm p},t;{\bm r}',t')\Big)+\mathcal{O}(\lambda^2).
\label{magnus7}
\end{equation}
By comparing Eqs.~\eqref{magnus5} and~\eqref{magnus7}, we conclude that $\exp{\big[\ii F^{(0)}({\bm p},t;{\bm r}',t')\big]}=\tilde K^{(0)} ({\bm p},t;{\bm r}',t')$
and $\tilde K^{(0)} ({\bm p},t;{\bm r}',t') F^{(1)}({\bm p},t;{\bm r}',t')=\tilde K^{(1)}_{\rm B} ({\bm p},t;{\bm r}',t')$. Hence, by performing the Born and Magnus 
expansions of the transformed propagator simultaneously, the terms $F^{(\ell)}({\bm p},t;{\bm r}',t')$ can be determined. However, it has to be noted that the complexity 
of the procedure increases with $\ell$, as for higher orders in $\lambda$ a larger number of factors, arising from the series expansion in~\eqref{magnus6}, contribute to the particular term.

Up to now our analysis has been very general. In the following, the Hamiltonian $H({\bm r}',t')$ will describe an electron interacting with an oscillating laser field, which 
in the dipole approximation is defined by the vector potential ${\bm A}(t)$, and with a general time-dependent binding potential, $V({\bm r},t)$, in the Kramers-Henneberger frame.

\section{Magnus and Born expansions in the Kramers-Henneberger frame}
Our aim is to derive an approximate expression for the propagator $\tilde K ({\bm p},t;{\bm r}',t')$ in the Kramers-Henneberger (KH) frame. Such frame is particularly suitable to perform our calculations, as the interaction with the scalar and vector potentials is reduced to a single term in the Hamiltonian.  

The retarded Schr\"odinger propagator in the Kramers-Henneberger (or accelerating) frame~\cite{kh1,kh2}, here denoted as $\krkh({\bm r},t;{\bm r}',t')$, satisfies the differential equation [see, Eq.~\eqref{general1}]
\begin{align}
\Big[-\ii\partial_{t'}&-H_{\rm KH}({\bm r}',t')\Big]\krkh({\bm r},t;{\bm r}',t')=\ii\delta(t-t')\delta({\bm r}-\bm{r}'),
\label{gauge3}
\end{align}
where $H_{\rm KH}({\bm r},t)=H_0+V({\bm r}+{\bm \alpha}(t),t)$ and $H_0=-\Delta/2\me$ is the free particle Hamiltonian.
${\bm \alpha}(t)$ is the so-called {\it displacement} vector, and it relates to the oscillating electric field ${\bm{\mathcal{E}}}(\tau)=-\partial_t{\bm A}(\tau)$ such that $\me\ddot{\bm \alpha}(\tau)=-e\dot{\bm A}(\tau)=e{\bm{\mathcal {E}}}(\tau)$. This, in turn, suggests that ${\bm \alpha}(\tau)$ determines the classical trajectory of a free electron in the light field, and $\ddot{\bm \alpha}(\tau)$ corresponds to its acceleration~\cite{kh2}. In this paper, it is assumed that the laser pulse acts over a finite period of time (from $t'$ up to $t$), which means that ${\bm {\mathcal E}}(\tau)={\bm 0}$ for $\tau\leqslant t'$ and $\tau \geqslant t$. We also choose the vector potential such that ${\bm { A}}(t)={\bm { A}}(t')={\bm 0}$, and assume that the displacement vanishes at times $\tau<t'$.

We proceed with expanding the propagator in a Born series. This is done by using repeatedly the Lippmann-Schwinger equation,
\begin{align}
\krkh({\bm r},t;{\bm r}',t')=\krkho({\bm r},t;{\bm r}',t')-\ii\int \dd^3y\,\dd\tau \krkho({\bm r},t;{\bm y},\tau)V({\bm y}+{\bm \alpha}(\tau),\tau)\krkh({\bm y},\tau;{\bm r}',t'),
\label{Lippmann-Schwinger}
\end{align}
where $\krkho({\bm r},t;{\bm r}',t')$ is the free-particle propagator. In the first-order Born approximation, we obtain that
\begin{align}
K^{(1)}_{\rm KH}({\bm r},t;{\bm r}',t')\approx\krkho({\bm r},t;{\bm r}',t')-\ii\int \dd^3y\,\dd\tau \krkho({\bm r},t;{\bm y},\tau)V({\bm y}+{\bm \alpha}(\tau),\tau)\krkho({\bm y},\tau;{\bm r}',t').
\label{gauge8}
\end{align}
Note that such general iterative procedure leads to a series expansion of the total propagator in powers of the binding potential. This is the essence of the BSSFA.

Since we are interested in deriving an expression for the propagator in momentum space, we perform the Fourier transform of $K^{(1)}_{\rm KH}({\bm r},t;{\bm r}',t')$ 
in~\eqref{gauge8} with respect to ${\bm r}$. In addition, since our main focus is the photoionization of atoms/ions, the binding potential 
is static. This means that $V({\bm y}+{\bm \alpha}(\tau),\tau)\equiv V({\bm y}+{\bm \alpha}(\tau))$ and we can use
\begin{equation}
V({\bm y}+{\bm \alpha}(\tau))=\int\frac{\dd^3k}{(2\pi)^3}\tilde V({\bm k})\ee^{\ii{\bm k}\cdot({\bm y}+{\bm \alpha}(\tau))}.
\label{prop24}
\end{equation}
As a result,
\begin{equation}
\tilde K^{(1)}_{\rm KH}({\bm p},t;{\bm r}',t')=\tkrkho({\bm p},t;{\bm r}',t')\bigg[1-\ii\!\!\int_{t'}^t\!\!\dd\tau\frac{d^3k}{(2\pi)^3}\tilde V({\bm k})
\exp{\Big(-\ii\frac{{\bm k}^2}{2\me}(\tau-t')+\ii{\bm k}\cdot{\bm R}_{\bm p}({\bm r}',\tau)}\Big)\bigg],
\label{prop31}
\end{equation}
where one can derive that 
$\tkrkho({\bm p},t;{\bm r}',t')=\exp{\big[-\ii{{\bm p}^2}(t-t')/2\me-\ii{\bm p}\cdot{\bm r}'\big]}$. 
Moreover, in Eq.~\eqref{prop31}, we have introduced a classical trajectory of an electron in the laser field, ${\bm R}_{\bm p}({\bm r}',\tau)={\bm r}'+{\bm \alpha}(\tau)+{\bm p}{}(\tau-t')/\me$.
Note that the second term in the square bracket in Eq.~\eqref{prop31} can be related to an {\it effective potential} experienced by the electron,
\begin{equation}
V_{\rm eff}({\bm R},\tau-t')=\int\frac{d^3k}{(2\pi)^3}\tilde V(\bm k)\exp{\Big(-\ii\frac{{\bm k}^2}{2\me}(\tau-t')+\ii{\bm k}\cdot{\bm R}}\Big).
\label{GEA1}
\end{equation}
Having this in mind, we rewrite~\eqref{prop31} in a more compact form,
\begin{align}
\tilde K^{(1)}_{\rm KH}({\bm p},t;{\bm r}',t')=&\tkrkho({\bm p},t;{\bm r}',t')\Big[1-\ii\int_{t'}^t\dd\tau V_{\rm eff}({\bm R}_{\bm p}({\bm r}',\tau),\tau-t')\Big],
\label{GEA2}
\end{align}
which represents the Magnus expansion of the propagator in momentum space~\eqref{magnus7} with 
\begin{equation}
\tkrkho({\bm p},t;{\bm r}',t')\equiv\exp{\big[\ii F^{(0)}({\bm p},t;{\bm r}',t')\big]}\quad \text{and}\quad F^{(1)}({\bm p},t;{\bm r}',t')\equiv-\int_{t'}^t\dd\tau V_{\rm eff}({\bm R}_{\bm p}({\bm r}',\tau),\tau-t').
\label{GEA3}
\end{equation}
It allows us to formally write down that, up to the linear term in the potential, 
\begin{equation}
\tilde K^{(1)}_{\rm KH}({\bm p},t;{\bm r}',t')\approx\tkrkho({\bm p},t;{\bm r}',t')\exp{\Big[-\ii\int_{t'}^t\dd\tau V_{\rm eff}({\bm R}_{\bm p}({\bm r}',\tau),\tau-t')\Big]}.
\label{GEA4}
\end{equation}
This expression is fundamental for our further analysis. As we will demonstrate shortly, the fact that the propagator depends on an effective potential rather than on the classical one, has very important consequences in the photoionization dynamics.

\subsection{Effective CCSFA}
In order to define our effective CCSFA, let us modify the classical trajectory of the electron in the laser field ${\bm R}_{\bm p}({\bm r}',\tau)$ by a correction 
that is linear in $\lambda$. Namely, we introduce an effective trajectory ${\bm q}(\tau)={\bm R}_{\bm p}({\bm r}',\tau)+\lambda\delta{\bm q}(\tau)$
which satisfies the same boundary conditions as the classical trajecotry; namely, ${\bm q}(t')={\bm r}'$ and $\me\dot{\bm q}(t)={\bm p}$. By making a transformation from the KH to the laboratory frame, 
we obtain that, up to linear terms in $\lambda$, the propagator~\eqref{GEA4} is given by
\begin{align}
\tsprvg^{(1)}({\bm p},t;{\bm r}',t')=&\exp\bigg[-\ii\int_{t'}^t\dd\tau\Big(\frac{\me}{2}\dot {\bm q}^2(\tau)+V_{\rm eff}({\bm q}(\tau),\tau-t')\Big)-\ii{\bm p}\cdot{\bm r'}\bigg]\\
=&\exp\big[\ii S_{\rm eff}[{\bm r},t;{\bm r}',t'|{\bm q}]-\ii\me\dot{\bm q}(t)\cdot{\bm q}(t)\big],
\label{GEA6}
\end{align}
where the subscript $V$ stands for the velocity gauge and where ${\bm r}={\bm q}(t)$, for fixed $t'$ and $t$, is a function of ${\bm r}'$ and ${\bm p}$. The functional $S_{\rm eff}[{\bm r},t;{\bm r}',t'|{\bm q}]$ in~\eqref{GEA6} 
is given by
\begin{equation}
S_{\rm eff}[{\bm r},t;{\bm r}',t'|{\bm q}]=\int_{t'}^t\dd\tau\Big(\frac{\me}{2}\dot {\bm q}^2(\tau)-V_{\rm eff}({\bm q}(\tau),\tau-t')+e{\bm{A}}(\tau)\cdot\dot{\bm q}(\tau)\Big),
\label{GEA7}
\end{equation}
which can be recognized as the classical action of an electron interacting with the laser field and the scalar effective potential $V_{\rm eff}({\bm r},\tau-t')$. 
This, in turn, implies that the dynamics of the system is governed by the effective (and complex) Newton equation
\begin{equation}
\me \ddot{\bm q}(\tau)=-{\bm \nabla}V_{\rm eff}({\bm q}(\tau),\tau-t')+e{\bm {\mathcal E}}(\tau).
\label{GEA8}
\end{equation}
$V_{\rm eff}({\bm q}(\tau),\tau-t')$, being a complex function, differs from the classical potential $V({\bm r})$ as it contains important quantum corrections 
(see, Sec.~\ref{classical_limit}). Furthermore, the effective potential~\eqref{GEA1} depends explicitly on the initial time $t'$, meaning that the Newton 
equation~\eqref{GEA8} contains information about past events. In other words, contrary to its classical counterpart, the effective quantum-mechanical 
evolution is non-Markovian.

Now, we present $V_{\rm eff}({\bm q}(\tau),\tau-t')$ for the spherically-symmetric Coulomb potential $V({\bm r})\equiv V(r)=-{Z\alpha c}/{r}$. Here, $\alpha$ is the fine-structure constant, $Z$ is the atomic number, and $c$ is the speed of light. From the definition~\eqref{GEA1}, it follows that (see, Refs. \cite{KK2015,VKK2016})
\begin{align}
V_{\rm eff}({\bm q},\tau-t')=-\frac{Z\alpha c}{q}{\rm erf}\bigg[\ee^{-\ii\pi/4}\sqrt{\frac{\me{\bm q}^2}{2(\tau-t')}}\bigg]=-Z\alpha c\sqrt{\frac{\me}{2(\tau-t')}}\,\frac{{\rm erf}[\ee^{-\ii\pi/4}\rho_S]}{\rho_S},
\label{potential5}
\end{align}
where $\rho^2_S=\me{\bm q}^2/{2(\tau-t')}$. The mathematical and physical consequences of this expression are analyzed below.

\section{Classical limit of the effective CCSFA and its range of validity}
\label{classical_limit}

The propagator in the effective CCSFA [Eq.~\eqref{GEA4}] depends on the effective potential~\eqref{potential5}, which differs from the pure Coulomb one. 
As a result, the effective interaction already accounts for the spreading of the electron wave packet and quantum diffusion effects~\cite{KK2015,VKK2016}. 
Such effects, while negligible at infinitesimally-short time intervals, become relevant at the characteristic time-scale at which photoionization takes 
place (see below). In addition, $V_{\rm eff}({\bm q},\tau-t')$ is an entire function of ${\bm q}^2$, i.e., it does not have singularities or branch points 
when the electron trajectories come back to the parent ion. This is in contrast to the pure Coulomb potential, as it is singular at the origin of coordinates. 
To see this better, consider the power-series expansion of the error function (see, Ref.~\cite{olver2010}) in Eq.~\eqref{potential5},
\begin{equation}
{\rm erf}(z)=\frac{2}{\sqrt{\pi}}\sum_{n=0}^\infty\frac{(-1)^nz^{2n+1}}{n!(2n+1)}.
\label{properties1}
\end{equation} 
Thus, the effective potential can be expressed as an absolutely convergent power series in the whole space,
\begin{equation}
V_{\rm eff}({\bm q},\tau-t')=-2Z\alpha c\ee^{-\ii\pi/4}\sqrt{\frac{\me}{2\pi(\tau-t')}}\sum_{n=0}^\infty\frac{1}{n!(2n+1)}\bigg(\ii\frac{\me {\bm q}^2}{2(\tau-t')}\bigg)^n,
\label{properties2}
\end{equation}
and, in particular, $\displaystyle\lim_{|{\bm q}|\to 0}|V_{\rm eff}({\bm q})|=2Z\alpha c\sqrt{{\me}/{2\pi(\tau-t')}}<\infty$,
provided that $\tau>t'$. On the other hand, for large arguments, the effective potential can be analyzed by making use of the asymptotic expansion of the 
complementary error function, ${\rm erfc}(z)=1-{\rm erf}(z)$ (see, Ref.~\cite{olver2010}). In doing so, we find that for large $|{\bm q}|$ 
(or, equivalently, for large $\rho_S$), $V_{\rm eff}({\bm q})\sim -{Z\alpha c}/{|{\bm q}|}$, i.e., the Coulomb interaction is recovered when the argument of the effective potential is large enough. 

\begin{figure}
\centering
\includegraphics[width=0.8\textwidth]{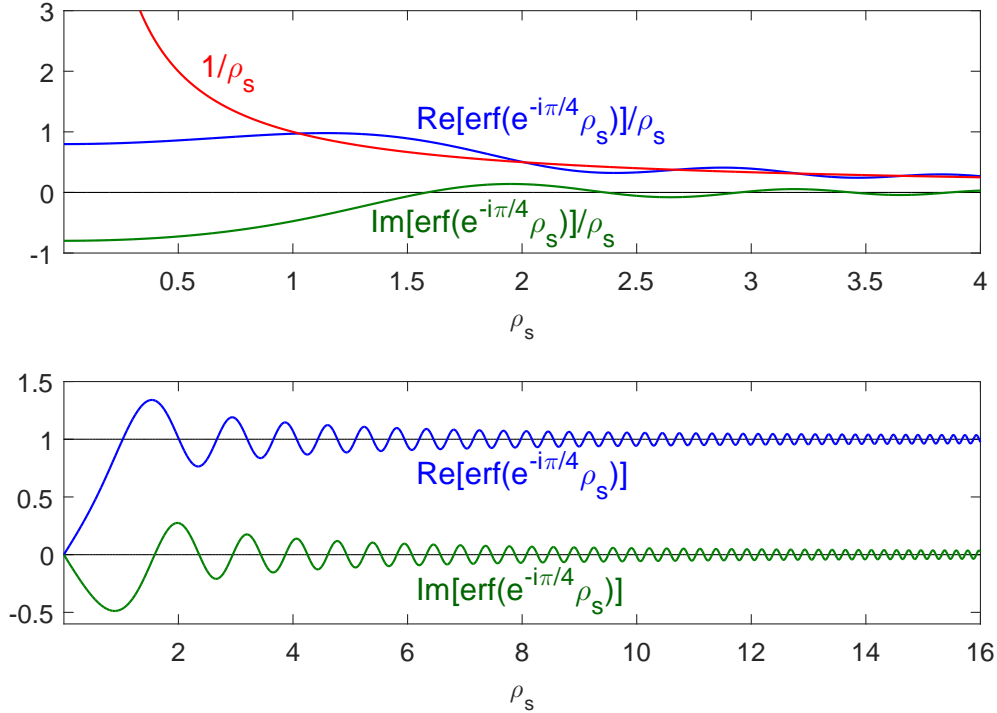}
\caption{In the upper panel we show the real (blue line) and imaginary (green line) parts of the function ${\rm erf }(\ee^{-\ii\pi/4}\rho_S)/\rho_S$, which determines the behavior of the effective potential~\eqref{potential5}. The red curve corresponds to $1/\rho_S$, i.e., it defines the Coulomb potential.  It is shown that for $\rho_S > 1$ the effective interaction of electrons with their parent ions is quite well described by the classical potential. However, when their trajectories come back to the origin of coordinates ($\rho_S <1$) such assumption is not valid anymore. In this case, the effective interaction, which is free of the problems related to the Coulomb singularity and branch points, has to be accounted for. In the lower panel of this figure we present the real (blue line) and imaginary (green line) parts of ${\rm erf }(\ee^{-\ii\pi/4}\rho_S)$. Note that the approximation ${\rm erf }(\ee^{-\ii\pi/4}\rho_S)\approx1$ is quite well satisfied for $\rho_S\gtrsim 1$.
\label{CoulombEffective}}
\end{figure}

In the upper panel of Fig.~\ref{CoulombEffective} we show the real (blue line) and imaginary (green line) parts of ${\rm erf }(\ee^{-\ii\pi/4}\rho_S)/\rho_S$ 
together with the curve $1/\rho_S$ (red line). While the first function determines the behavior of the effective potential [Eq.~\eqref{potential5}], the second one 
defines the classical Coulomb interaction. It can be seen that the imaginary part of $V_{\rm eff}$ goes fast to zero for large values of $\rho_S$. In contrast, 
${\rm Re}[{\rm erf }(\ee^{-\ii\pi/4}\rho_S)]/\rho_S$ behaves in a very similar way as $1/\rho_S$, when $\rho_S>1$. This is in agreement with our previous analysis 
which established that, in the limit of large arguments, the effective potential and the classical one coincide. On the other hand, for small values of $\rho_S$ 
the situation changes drastically. First, ${\rm Im}[{\rm erf }(\ee^{-\ii\pi/4}\rho_S)]/\rho_S$ is nonzero and remains finite. Second, the general behavior 
of ${\rm Re}[{\rm erf }(\ee^{-\ii\pi/4}\rho_S)]/\rho_S$ and $1/\rho_S$ differ considerably: while the former remains bounded and reaches values near to one,
the latter increases rapidly and presents a singularity at $\rho_S=0$. This agrees with our earlier conclusions that the effective potential is an entire 
function of $\rho^2_S$ and does not contain singularities or branch points. The dramatic differences in the behavior of the two potentials near the origin of 
coordinates (${\bm q}\approx{\bm 0}$) can be attributed to the inherent quantum-mechanical nature of photoionization; diffusion and spreading of the wave packet 
play an important role in the process and they are neglected in a classical treatment or CCSFA.

\begin{figure}
\centering
\includegraphics[width=0.65\textwidth]{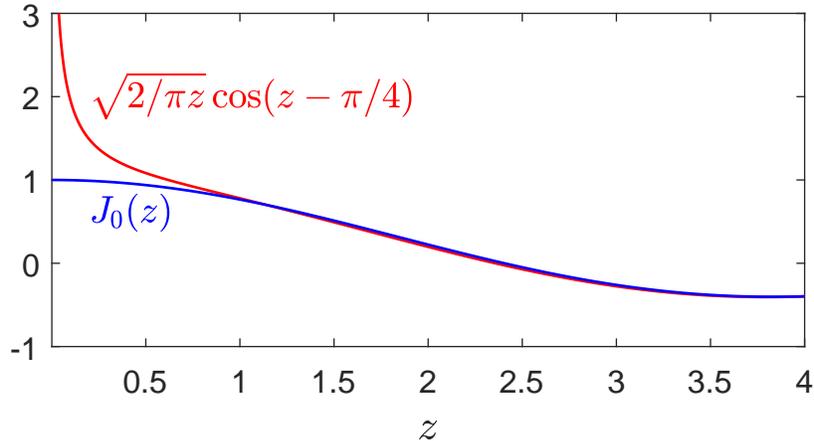}
\caption{Comparison of the Bessel function $J_0(z)$ with its asymptotic expansion. 
\label{BesselJ0}}
\end{figure}

As we have shown, {\it only} the asymptotic form of $V_{\rm eff}({\bm q},\tau-t')$ (which is valid provided that $|{\bm q}|$ is large) has singularities 
and branch points at small values of ${\bm q}$. A very similar situation is met in the theory of special functions. As an example, consider the Bessel 
function $J_0(z)$, which is an entire function of $z^2$ (cf., Fig.~\ref{BesselJ0}). Its power series expansion,
\begin{equation}
J_0(z)=\sum_{n=0}^\infty (-1)^n\frac{(z^2/4)^n}{(n!)^2},
\label{properties6}
\end{equation}
is absolutely convergent and well-behaved for all complex arguments $z$. However, its asymptotic behavior, $J_0(z)\sim \sqrt{{2}/{(\pi z)}}\cos(z-{\pi}/{4})$ 
for $|{\rm arg}(z)|<\pi$, might indicate that $J_0(z)$ exhibits the singularity and the branch point at $z=0$. This is not the case because the asymptotic form is not applicable there. We illustrate this in Fig.~\ref{BesselJ0}, where $J_0(z)$ (blue line) and its asymptotic expansion (red line) are presented.

Now, an important question arises: What are the arguments of $V_{\rm eff}({\bm q},\tau-t')$  for which the effective potential can be well approximated by its asymptotic (classical) form? This can happen for real $\rho_S$ such that ${\rm erf}(\ee^{-\ii\pi/4}\rho_S)\approx1$. In the lower panel of Fig.~\ref{CoulombEffective} we present the real and imaginary parts of this function (blue and green lines, respectively). 
Hence, one can estimate that the condition ${\rm erf}(\ee^{-\ii\pi/4}\rho_S)\approx1$ is quite well fulfilled for $\rho_S\gtrsim 1$, i.e., the classical form of the effective potential can be applied provided that
\begin{equation}
\frac{\me }{2(\tau-t')}{\bm q}^2\gtrsim1.
\label{properties10}
\end{equation}
This condition is satisfied if either:
\begin{enumerate}
\item\label{case1} For fixed $\me$ and ${\bm q}^2$, the evolution happens in a very short time interval ($\tau\approx t'$). This is the basic assumption which leads to the Feynman path integral representation of the propagator~\cite{path1,path2,path3}.
\item\label{case2} For fixed $(\tau-t')$ and ${\bm q}^2$, the particle mass is very large. In fact, such statement is the essence of the Born-Oppenheimer approximation, i.e., it is assumed that the dynamics of massive particles can be approximated by the classical treatment.
\item\label{case3} For fixed $\me$ and $(\tau-t')$, the trajectories do not come back close to the origin of coordinates. This limit is justified, for instance, in the ionization of Rydberg states (RS) by microwave fields. If the distance between the electron and the nucleus is approximately equal to the Bohr radius of the RS (with principal quantum number $n\gg1$), and $\tau-t'$ corresponds to its orbital period, the classical limit is applicable provided that $n/(4\pi)\gtrsim 1$ [see, Eq.~\eqref{properties10}]. Thus, for $n\gtrsim 12$ the electron dynamics is determined by the classical evolution~\cite{classrf1,classrf2}.
\item\label{case4} 
In scattering processes the condition~\eqref{properties10} is equivalent to $|{\bm p}| \cdot{b}\gtrsim 1$, where ${\bm p}$ is the electron momentum and $b$ is the impact parameter. This is, in fact, the condition of applicability of the traditionally-used eikonal approximation. Hence, such approximation is valid provided that $|{\bm p}|$ and/or $b$ are sufficiently large and it excludes the case of backward scattering.
\end{enumerate}
It follows from here that the asymptotic expansion of the effective potential can be applied, for instance, to ionization driven 
by elliptically polarized laser fields, as the electron trajectories do not return to the parent ion. The situation is different when a linearly
polarized pulse is considered. In such case, the electron can come back to the origin of coordinates at times close to half of the laser cycle, i.e., 
$\tau-t'\sim{\pi}/{\omega_{\rm L}}$, where $\omega_{\rm L}$ is the carrier frequency of the laser field. This means that the asymptotic limit 
(i.e., the classical theory or the CCSFA) is {\it only} justified provided that ${\bm q}^2\gtrsim{2\pi}/{\omega_{\rm L}}$, in atomic units. 
For a typical frequency of $\omega_{\rm L}=1.5$~eV$\approx1/20$~a.u., we have ${\bm q}^2\gtrsim 40\pi$ or, equivalently, $|{\bm q}|\gtrsim12$~a.u. 
This means that the electron trajectory should be far away from the parent ion. The range of applicability of the classical description of ionization is improved
for X-ray laser pulses. For instance, for $\omega_{\rm L}\approx 100$~eV the electron can approach the origin of coordinates up to a distance 
of the order of $a_0$, where $a_0$ is the Bohr radius. 

\section{Conclusions}

Both CCSFA and the classical approach are asymptotic limits of the quantum theory based on the Schr\"odinger equation. Therefore, they can be used only within 
their domains of validity. We suggest that, beyond those domains, either the methods based on the Born expansion or the effective CCSFA with the regular complex effective action should be used. 
Our analysis has been limited to the case when the effective action is a linear functional of the binding potential. However, by considering higher terms in the Born expansion or the eikonal perturbation theory, one can derive further modifications of the effective CCSFA which are consistent with the quantum theory.

\section*{Acknowledgments}
This work is supported by the National Science Centre (Poland) under Grant No. 2014/15/B/ST2/02203.


\section*{References}
\begin{thebibliography}{99}

\bibitem{BK} 
Brabec T and Krausz F 2000 {\it Rev. Mod. Phys.} {\bf 72} 545

\bibitem{IK} 
Krausz F and Ivanov M 2009 {\it Rev. Mod. Phys.} {\bf 81} 163

\bibitem{Keldysh}
Keldysh L V 1964 {\it Zh. Eksp. Teor. Fiz.} {\bf 47} 1945; 1965 {\it Sov. Phys. JETP} {\bf 20} 1307

\bibitem{Faisal}
Faisal F H M 1973 {\it J. Phys.} B {\bf 6} L89

\bibitem{Reiss}
Reiss H R 1980 {\it Phys. Rev.} A {\bf 22} 1786

\bibitem{kh1}
Kramers H A 1956 {\it Collected Scientific Papers} (Amsterdam: North-Holland) p 262

\bibitem{kh2}
Henneberger W C 1968 {\it Phys. Rev. Lett.} {\bf 21} 838

\bibitem{Volkov}
Wolkow D V 1935 {\it Z. Phys.} {\bf 94} 250

\bibitem{EJK1998} 
Ehlotzky F, Jaro\'n A and Kami\'nski J Z 1998 {\it Phys. Rep.} {\bf 297} 63

\bibitem{Fedorov1}
Bunkin F V and Fedorov M V 1965 {\it Zh. Eksp. Teor. Fiz.} {\bf 49} 1215; 1966 {\it Sov. Phys. JETP} {\bf 22} 844

\bibitem{Fedorov2}
Denisov M M and Fedorov M V 1967 {\it Zh. Eksp. Teor. Fiz.} {\bf 53} 1340; 1968 {\it Sov. Phys. JETP} {\bf 26} 779

\bibitem{EKK2009} 
Ehlotzky F, Krajewska K and Kami\'nski J Z 2009 {\it Rep. Prog. Phys.} {\bf 72} 046401

\bibitem{DiPiazza2012} 
Di Piazza A, M\"uller C, Hatsagortsyan K Z and Keitel C H 2012 {\it Rev. Mod. Phys.} {\bf 84} 1177

\bibitem{DiPiazza2018} 
Di Piazza A 2018 {\it Phys. Rev.} D {\bf 97} 056028

\bibitem{Tsoar}
Jain M and Tzoar N 1978 {\it Phys. Rev.} A {\bf 18} 538

\bibitem{Kornev}
Kornev A S and Zon B A 2002 {\it J. Phys. B: At. Mol. Opt. Phys.} {\bf 35} 2451

\bibitem{Duchateau}

Duchateau G 2018 {\it Eur. Phys. J. Plus} {\bf 133} 186

\bibitem{Ferrante}
Basile S, Trombetta F and Ferrante G 1988 {\it Phys. Rev. Lett.} {\bf 61} 2435

\bibitem{Milo2018}
Gazibegovi\'c-Busulad\v{z}i\'c A, Becker W and Milo\v{s}evi\'c D B 2018 {\it Opt. Express} {\bf 26} 12684

\bibitem{K1}
Kami\'nski J Z 1986 {\it Phys. Scr.} {\bf 34} 770

\bibitem{K2}
Kami\'nski J Z, Jaro\'n A and Ehlotzky F 1996 {\it Phys. Rev.} A {\bf 53} 1756

\bibitem{PU2018}
Pyak P E and Udachenko V I 2018 {\it J. Phys. B: At. Mol. Opt. Phys.} {\bf 51} 065001 

\bibitem{ZZ2018}
Zaytzev A S, Zaytzev S A, Ancarani L U and Kouzakov K A 2018 {\it Phys. Rev.} A {\bf 97} 043417

\bibitem{Gav1}
Offerhaus M J, Kami\'nski J Z and Gavrila M 1985 {\it Phys. Lett.} A {\bf 112} 151

\bibitem{Gav2}
Gavrila M 2002 {\it J. Phys. B: At. Mol. Opt. Phys.} {\bf 35} R147


\bibitem{Resc1}
Becker W, Lohr A and Kleber M 1995 {\it Quantum Semiclass. Opt.} {\bf 7} 423

\bibitem{eik1}
Choudhury B J and Bakar B S 1974 {\it J. Phys.} B {\bf 7} L137

\bibitem{eik2}
Choudhury B J and Bakar B S 1975 {\it J. Phys.} B {\bf 8} L228

\bibitem{eik3}
Zon B A 1975 {\it J. Phys.} B {\bf 8} L86

\bibitem{eik4}
Kami\'nski J Z 1984 {\it Acta Phys. Pol.} A {\bf 66} 517

\bibitem{KK2015} 
Cajiao V\'{e}lez F, Krajewska K and Kami\'{n}ski J Z 2015 {\it Phys. Rev.} A {\bf 91} 053417

\bibitem{VKK2016}
Cajiao V\'{e}lez F, Krajewska K and Kami\'{n}ski J Z 2016 {\it J. Phys.: Conf. Ser.} {\bf 691} 012003

\bibitem{ccsfa1}
Popruzhenko S V, Mur V D, Popov V S and Bauer D 2008 {\it Phys. Rev. Lett.} {\bf 101} 193003

\bibitem{ccsfa2}
Pisanty E and Ivanov M 2016 {\it Phys. Rev.} A {\bf 93} 043408

\bibitem{ccsfa3}
Maxwell A S, Al-Jawahiry A, Das T and Figueira de Morisson Faria C 2017 {\it Phys. Rev.} A {\bf 96} 023420

\bibitem{ccsfa4}
Shvetsov-Shilovski N I and Lein M 2018 {\it Phys. Rev.} A {\bf 97} 013411

\bibitem{classdyn}
Grozdanov T, Gruji\'c P and Krsti\'c P 1989 {\it Classical Dynamics in Atomic and Molecular Physics} (Singapore: World Scientific)

\bibitem{classion1}
Gajda M, Grochmalicki J, Lewenstein M and Rz\c a\.zewski K 1992 {\it Phys. Rev.} A {\bf 46} 1638

\bibitem{classion2}
Schmitz H, Boucke K and Kull H J 1998 {\it Phys. Rev.} A {\bf 57} 467

\bibitem{classion2a}
El-Khawaldeh A and Kull H J 2017 {\it Phys. Rev.} A {\bf 95} 043401

\bibitem{classresc}
Mauger F, Chandre C and Uzer T 2010 {\it Phys. Rev. Lett.} {\bf 105} 083002

\bibitem{classrf1}
Leopold J G and Parsival I C 1978 {\it Phys. Rev. Lett.} {\bf 41} 944

\bibitem{classrf2}
Leopold J G and Parsival I C 1979 {\it J. Phys.} B {\bf 12} 709

\bibitem{Nurhuda}
Nurhuda M 2004 {\it Comp. Phys. Commun.} {\bf 162} 1


\bibitem{num1}
Majorosi S and Czirj\'ak A 2016 {\it Comp. Phys. Commun.} {\bf 208} 9

\bibitem{press0}
Bethune-Waddell M and Chau K J 2015 {\it Rep. Prog. Phys.} {\bf 78} 122401

\bibitem{KKpress}
Krajewska K and Kami\'nski J Z 2015 {\it Phys. Rev.} A {\bf 92} 043419

\bibitem{press1}
Smeenk C T L, Arissian L, Zhou B, Mysyrowicz A, Villeneuve D M, Staudte A and Corkum P B 2011 {\it Phys. Rev. Lett.} {\bf 106} 193002

\bibitem{press2}
Titi A S and Drake G W F 2012 {\it Phys. Rev.} A {\bf 85} 041404(R)

\bibitem{press3}
Reiss H R 2013 {\it Phys. Rev.} A {\bf 87} 033421

\bibitem{press4}
Chelkowski S, Bandrauk A D and Corkum P B 2014 {\it Phys. Rev. Lett.} {\bf 113} 263005

\bibitem{press5}
Ivanov I A 2015 {\it Phys. Rev.} A {\bf 91} 043410

\bibitem{press6}
He P-L, Lao D and He F 2017 {\it Phys. Rev. Lett.} {\bf 118} 163203

\bibitem{CKKvortex}
Cajiao V\'{e}lez F, Krajewska K and Kami\'{n}ski J Z 2018 {\it Phys. Rev.} A {\bf 97} 043421

\bibitem{magnus1}
Magnus W 1954 {\it Comm. Pure Appl. Math.} {\bf 7} 649

\bibitem{magnus2}
Mananga E S and Charpentier T 2016 {\it Phys. Rep.} {\bf 609} 1

\bibitem{magnus2a}
Blanes S, Casas F, Oteo J A and Ros J 2009 {\it Phys. Rep.} {\bf 470} 151

\bibitem{magnus3}
Kuwahara T, Moria T and Saito K 2016 {\it Ann. Phys.} {\bf 367} 96



\bibitem{olver2010}
Olver F W J, Lozier D W, Boisvert R F and Clark C W 2010 {\it NIST Handbook of Mathematical Functions} (New York: Cambridge)

\bibitem{path1}
Dirac P A M 1933 {\it Phys. Z. Sowjetunion} {\bf 3} 64

\bibitem{path2}
Feynman R P 1948 {\it Rev. Mod. Phys.} {\bf 20} 367

\bibitem{path3}
Kleinert H 2006 {\it Path Integrals in Quantum Mechanics, Statistics, Polimer Physics, and Financial Markets} (Singapore: World Scientific)







\end{thebibliography}
\end{document}